\documentstyle[11pt,psfig]{article}
\begin{document}

\begin{flushright} 
UMDGR-97-110\\ 
hep-th/9705017\\ 
\end{flushright} 
\vskip 1cm 
\begin{center}
{\Huge Semiclassical Decay of Near-Extremal\\ \vskip 2mm Black Holes }\\ 
\vskip 1cm 
{\large Ted Jacobson} 
\vskip .5cm 
{\it Department of Physics, University of Maryland\\
College Park, MD 20742-4111, USA}\\
	     {\tt jacobson@physics.umd.edu}

\end{center} \vskip 1cm

\begin{abstract}
Decay of a near-extremal black hole down to the extremal state is
studied in the background field approximation to determine the
fate of injected matter and Hawking pairs. By examining the behavior
of light rays and solutions to the wave equation it is concluded that
the singularity at the origin is irrelevant. Furthermore, 
there is most likely an instability of the event horizon arising 
from the accumulation of injected matter and Hawking partners there. 
The possible role of this instability in reconciling the D-brane and black
hole pictures of the decay process is discussed. 
 
\end{abstract}

\def\beq{\begin{equation}}
\def\eeq{\end{equation}}
\def\gsim{\; \raisebox{-.8ex}{$\stackrel{\textstyle >}{\sim}$}\;}
\def\lsim{\; \raisebox{-.8ex}{$\stackrel{\textstyle <}{\sim}$}\;}

\section{Introduction}
The decay of black holes by quantum emission of thermal
radiation suggests that quantum gravity is not unitary\cite{Hawk}.
Two types of information appear to be lost in the
decay process, the state of the matter that falls into the
black hole and the correlations between the Hawking quanta
that are radiated away and their ``partners" in the pair
creation process. If the black hole evaporates completely
this information seems to be destroyed, at least as far as
the outside observer is concerned.

The argument that quantum gravity is not unitary is of course
inconclusive at present since we lack a complete theory. According 
to one point of view the unitarity question hinges on the physics
of the curvature singularity inside the black hole. The information that
falls into the singularity might for example be destroyed, or leaked out
to the exterior in a non-local process, or it might be passed on to a baby
universe born at the singularity. Another viewpoint holds that it is 
not the singularity but the horizon that is the locus of hocus pocus.

The unitarity question has recently taken on a new light in view
of a string theoretic correspondence between extremal (and 
near-extremal) black holes and certain D-brane configurations in 
perturbative string theory. Calculations of entropy and
the rate of decay for these stringy states at weak coupling in
flat spacetime agree spectacularly with totally different calculations
for the corresponding black holes at strong coupling based on quantum 
field theory in a curved background black hole spacetime. 

How far does the agreement between the string and black hole
pictures go?  Does the presence of the event horizon and 
singularity inside the black hole make any difference to the
question of unitarity and the state of the outgoing radiation?
In this paper the semiclassical description
of the decay of a near-extremal black hole is studied 
as a first step in addressing these questions. 
The existence of a ground state makes this decay 
quite different from the evaporation of neutral
(or discharging) black holes.  

The model studied here is a spherically symmetric charged black 
hole that is excited above extremality and then allowed to decay 
back to the extremal ground state. This model was previously
studied analytically by Strominger and Trivedi\cite{StroTriv}
in the large-$N$, adiabatic, $S$-wave approximation 
(two-dimensional reduction), including the back-reaction at one-loop order.
Their work established the global structure of the spacetime 
in this approximation, from which they argued that an arbitrarily
large amount of information can be injected into the black hole 
and lost to the outside world. 
The same model was also studied numerically, without the large-$N$ 
or adiabatic approximations, by Lowe and O'Loughlin\cite{LoweO'Lo}, 
whose results lend support to this picture, although of course they 
only evolved the system for a finite time.
The focus of the present paper 
is in a sense complementary to the analyses of 
\cite{StroTriv} and \cite{LoweO'Lo}. 
I do not implement a self-consistent model of the back-reaction; 
however I analyze more closely what is happening to the quantum
fields inside the black hole. 

This analysis suggests some rather
surprising conclusions: First, the singularity at the origin is 
irrelevant. 
Second, the inner apparent horizon is probably 
quantum mechanically stable, unlike in the static case where
the locally measured energy density grows exponentially with 
time.
(If instead the inner horizon is unstable, 
then the semiclassical approximation
breaks down, and information can fall into the
strongly curved region.) 
If the inner horizon is indeed stable  
then the information in the black hole interior ends
up sitting just behind the event horizon, and the {\it event} horizon 
is quantum mechanically unstable.
It would  be very interesting to see whether the numerical approach
taken in \cite{LoweO'Lo} could be pushed to late enough times to
check the picture of the instability arrived at here by adiabatic 
arguments.
A horizon instability means of course that the semi-classical
analysis is not as it stands self-consistent. 
In the Discussion section I will speculate on the role of this instability
in reconciling the D-brane and black hole pictures, and 
in particular its relation to 
the singular horizons that occur in the spacetimes corresponding
to D-brane configurations of non-maximal entropy.
 
\section{The Model}
\label{model}

The process I will consider is the following. Starting with
an extremal black hole (which has vanishing Hawking temperature),
with a  stable charge\footnote{In the string 
calculations the charge cannot be radiated away
because only very massive solitons carry the charge. 
This is why the black hole decays back to an extremal state 
rather than discharging and evaporating completely.},
some matter is thrown in raising the temperature above zero.
The black hole then emits Hawking radiation (in whatever quantum
fields are present) and decays back to 
the extremal state. To model the spacetime of this process 
I use the charged Bonner-Vaidya metric \cite{PlebStac,BonnVaid,SullIsra}
in four dimensions:
\beq
ds^2= f(r,v) \, dv^2 - 2\, dvdr - 
r^2(d\theta^2 + \sin^2\theta\, d\phi^2)
\label{Vaidya}
\eeq
with
\beq
f(r,v)=1-\frac{2M(v)}{r}+\frac{Q(v)^2}{r^2}.
\label{f}\eeq
(Here and below I use units with $G=c=\hbar=1$.) This line element
is a solution to the Einstein-Maxwell-fluid equations with an ingoing charged
null fluid.  The charge density is proportional to $\dot{Q}\equiv dQ/dv$ 
and the
energy-momentum tensor is
$T_{\mu\nu}=\rho \nabla_\mu v \nabla_\nu v + T^{\rm em}_{\mu\nu}$,
where $\rho=(4\pi r^2)^{-1}(\dot{M}-Q\dot{Q}/r)$ and
$T^{\rm em}_{\mu\nu}$ is the electromagnetic field stress tensor
for the radial electric field. 
If the mass $M$ and charge $Q$ are constant (\ref{Vaidya})
is just the Reissner-Nordstr\"om black hole in ingoing 
Eddington-Finkelstein coordinates, and if $M=Q$ it is extremal.
This is the non-compact part of one of the black holes used in the D-brane
calculations\cite{StroVafa,CallMald,MaldStro1,JKM}. It seems 
not unlikely that the essential ideas discussed 
here would carry over to all near-extremal black holes.
The Vaidya ($Q=0$) and Vaidya-Bonner metrics have been used in many 
previous studies of evaporating black holes (see for example 
\cite{HiscEvap,Kami,StroTriv} and references therein). 

The Bonnor-Vaidya metric is not in detail the correct solution for our
problem, since in the evaporation process the stress-energy tensor
for the quantum fields is not described by a purely ingoing null
flux of negative energy. Although an ingoing negative energy flux exists,
there is also an outgoing positive energy flux outside the horizon,
an outgoing negative energy flux inside, and vacuum polarization terms,
However, if the evaporation process is adiabatic until a large stress energy
develops, the behavior of the quantum fields on the black hole background 
should be reasonably well modeled by employing a sequence of 
Reissner-Nordstr\"om metrics with fixed charge and decreasing mass.
The arguments of this paper depend only on the adiabaticity.
In particular, 
the conclusion that a large stress energy develops somewhere is
reliable in this approximation, although the subsequent evolution
would need to be understood in a dynamically consistent manner.

Now let us fix $Q(v)=Q\gg 1$, and define $\mu(v)$ as the mass above
extremality,
\beq
M(v) = Q + \mu(v),
\eeq
so that 
\beq
f(r,v)= (1-\frac{Q}{r})^2 - \frac{2\mu(v)}{r}.
\eeq
When the black hole is absorbing mass $\mu(v)$ is increasing.
When the Hawking radiation is emitted there is a negative
energy flux into the black hole. To model this process I 
take $\mu(v)$ to be decreasing during that period. Thus
$\mu(v)$ starts out zero, grows to a maximum, and then shrinks
back to zero. The rate of decrease of $\mu$ is the luminosity,
$\dot{\mu}\sim -T_H^4 A$, where $T_H=\kappa/2\pi$ is the Hawking 
temperature and $\kappa$ is the surface gravity. For 
a static near-extremal black hole ($\mu\ll Q$) we have 
\beq
\kappa=\frac{1}{2}(df/dr)\vert_{f=0}\simeq \sqrt{2\mu/Q^3},
\label{kappa}
\eeq
so
$\dot{\mu}\sim -\mu^2/Q^4$,
which implies 
\beq
\mu(v)\sim Q^4/v.
\label{halflife}
\eeq
That is, $\mu$ decays back to zero as $v^{-1}$, decreasing
to half its initial value in a ``half-life" of order $Q^4/\mu$.
Note that the Hawking temperature is much lower than $\mu$
until $\mu$ decreases to something of order $1/Q^3$, so that 
the semiclassical treatment of Hawking radiation looks quite 
reasonable for almost all of the decay process. 
Also, it is known\cite{AndeHiscLora} that the zero temperature
static vacuum state for a massless scalar field with arbitrary
curvature coupling is regular on the event horizon of an extremal
four-dimensional black hole.\footnote{Numerical calculations 
in \cite{AndeHiscLora} showed that the stress tensor is regular,
but its derivatives were not examined. In the two-dimensional 
case\cite{Triv} the stress tensor blows up on the horizon, but   
when one loop corrections are incorporated self-consistently in 
a dilaton-gravity-matter model, the divergence is postponed to the
second derivative of the stress tensor.}
Thus the semiclassical treatment of the decay initially
appears justified.
We shall find later that a horizon
instability at late times calls this into question however.

The zeroes of the metric component $f(r,v)$ are located at
\beq
r_\pm=Q+\mu\pm\sqrt{2\mu Q+\mu^2}.
\label{roots}
\eeq
For constant $\mu$, $r_+$ is the event horizon and $r_-$ is the
inner horizon or Cauchy horizon inside the Reissner-Nordstr\"om
black hole\cite{HawkElli}. Between $r_-$ and $r_+$ the
function $f$, which is the norm of the ``time" translation 
Killing field $\partial/\partial v$, is negative. This is the
ergoregion. It is also the region of outer trapped surfaces.
The ingoing radial light rays satisfy $dv=0$, while the ``outgoing"
ones are given by $f\, dv = 2dr$. Thus where $f<0$ the outgoing
radial light rays are in fact going to smaller values of $r$.
In the extremal case $\mu=0$, the two horizons coincide, and there
is no ergoregion. This is why there is no Hawking radiation:
no negative energy states are available for the partners of the
Hawking quanta. 

\begin{figure}[bt]
\centerline{
\psfig{figure=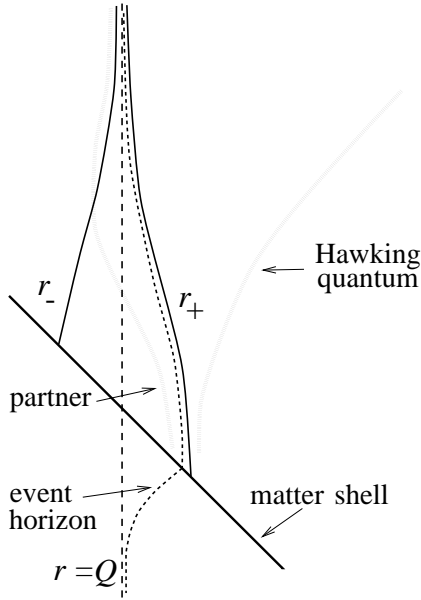,angle=0,height=8cm}}
\caption{\small Decay of a near-extremal black hole in Eddington-Finkelstein
coordinates $(v,r)$. Lines of constant $r$ are vertical and those
of constant $v$ slope at 45$^\circ$ towards the upper left.
The curves $r_\pm(v)$ are the outer and inner apparent horizons.
Hawking production of a pair of localized wavepackets is sketched.}
\label{decayfig}
\end{figure}

In the dynamical case, where $\mu(v)$ grows and then
shrinks back to zero, the two zeroes of $f$ split and then come
back together (see Fig. \ref{decayfig}). 
The region where $f$ is negative forms a blister
on the extremal horizon containing trapped surfaces. I call the
outer boundary of this region at $r_+$ the {\it outer apparent horizon}
(or sometimes just {\it apparent horizon}),
and the inner boundary at $r_-$ the {\it inner apparent horizon}
(or sometimes just {\it inner horizon}). 
The width in $r$ of the trapped region at constant $v$ is
$\Delta r:=r_+-r_-$. In the near-extremal
case (\ref{roots}) yields
\beq
\Delta r\simeq \sqrt{8\mu Q}.
\eeq
When the event horizon enters the trapped region  
it begins shrinking monotonically as the black hole loses mass.   
It must remain inside the trapped region (where $f<0$) from that point
on until the trapped region goes away (unless more positive energy matter is
thrown in at a later time).  
Note that at the boundary of the trapped region $f=0$, so the 
outgoing light rays must be `vertical' there, i.e., $dr/dv=0$.

If the extremal black hole is excited by a macroscopic
amount to a near-extremal state,
then it will remain for a long time as an approximate 
Reissner-Nordstr\"om black hole, slowly decaying back
to extremality. This is illustrated in Fig. \ref{decayfig}, in which
the injected mass is taken to be a null shell. 
Since the outgoing light rays are vertical wherever $f=0$, 
while the inner and outer apparent horizons are moving out and
in respectively,
the inner horizon is evidently spacelike while the outer (apparent)
horizon is timelike.
The event horizon remains very close to the apparent horizon
throughout the decay process. One can estimate just how close by 
noting that the event horizon must stay outside $r=Q$ for a time
of order the initial half-life $Q^4/\mu$. Using the expression
given in the next section for the time for a light ray to peel away 
from near the horizon
of a static black hole, this implies that the initial radial coordinate
of the event horizon must be $r_+-\epsilon$, with 
\beq
\epsilon\sim \sqrt{\mu Q} \exp(-\sqrt{Q^5/\mu}),
\label{epsilon}
\eeq
a very small distance indeed.

\section{Decay of a near-extremal black hole}
\label{decay}

The essential point of this paper arises from the elementary 
observation that the behavior of ``outgoing" light rays inside
the black hole is qualitatively very different for an extremal
black hole than it is for a neutral or charged nonextremal one. 
The outgoing radial light
rays are obtained by integrating the equation
\beq
dr/dv = f/2.
\eeq
In the neutral (Schwarzschild)
black hole, these rays peel away from the event horizon and fall to
the (spacelike) singularity at $r=0$ in a time (with respect to 
the ingoing advanced time coordinate $v$) of order 
$r_+\ln(r_+/\epsilon)$, where $r_+-\epsilon$ is the initial radial
coordinate.
In the non-extremal charged black hole, these
rays peel away from the event horizon reaching the midpoint $(r_+ + r_-)/2$
in a time of order $[r_+^2/(r_+-r_-)]\ln[(r_+-r_-)/\epsilon]$.
Then, rather than falling into the (timelike)
singularity at $r=0$, they asymptotically approach the inner
horizon in a time that diverges as 
$[r_-^2/(r_+-r_-)]\ln[(r_+-r_-)/\epsilon]$, where now $r_-+\epsilon$
is the final radial coordinate.
In the extremal
case, on the other hand, the outgoing light rays {\it never}
peel away from the event horizon. Instead they {\it approach} 
the event horizon as $(r_+-r)\sim Q^2/v$. 

We should pause here to understand exactly in what sense the
outgoing light rays get ``close to the horizon". An invariant
description can be given by reference to the freely falling
observers that start at rest at infinity and fall across the 
horizon. As $v\rightarrow\infty$, 
the outgoing light ray inside the event horizon will be intercepted by these
observers an arbitrarily short proper time after they cross
the horizon.\footnote{This follows from the radial equation 
(\ref{radeqn}). With $L=0$, $E=1$, and $\eta=1$ (\ref{radeqn}) yields
$\dot{r}=-(1-f)^{1/2}$. Thus near the horizon one has 
$\dot{r}\simeq 1$, so that the radial coordinate passes at the
same rate as the proper time.} 
\begin{figure}[tb]
\centerline{
\psfig{figure=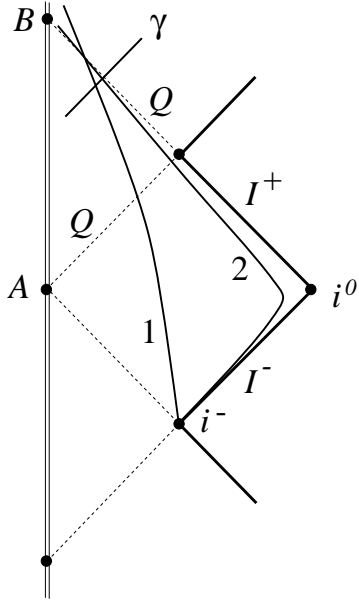,angle=0,height=8cm}}
\caption{\small Penrose diagram of part of the maximally extended extremal
Reissner-Nordstr\"om spacetime. The interior of the 
trapezoid $ABi^0i^-$ is the patch
covered by Eddington-Finkelstein coordinates. The dotted lines
labeled $Q$ are the event horizon and inner horizon. The 
two curves labeled $1$ and $2$ are radial free-fall trajectories
that begin at rest at infinity and fall into the black hole at 
different times. The proper time along the world line 2 between
the event horizon and the crossing of the outgoing light ray $\gamma$
goes to zero as the infall time of $2$ goes to infinity.}
\label{ern}
\end{figure}
It is worth pointing out that this is somewhat counter-intuitive 
from the point of view of the Penrose diagram for an extreme 
Reissner-Nordstr\"om metric, which is depicted in Fig. \ref{ern}. 
Both the inner horizon and the event horizon
occur at the radial coordinate $Q$. The Eddington-Finkelstein
coordinates cover only the interior of the trapezoidal region $ABi^0i^-$, 
so the inner horizon 
is not included in the Eddington-Finkelstein patch. 
The outgoing light ray $\gamma$ actually
crosses the inner horizon (in a finite affine parameter, although
the advanced time goes to infinity there). Nevertheless,
this ray also gets arbitrarily close to the event 
horizon, in the sense of the proper time of free-fall observers 
described above, even though on the
Penrose diagram it can appear to cross the
inner horizon very far from the event horizon.\footnote{Similar 
comments can be made about the non-extremal case as well.
There, the outgoing rays cross the ``top right" part of the inner
horizon at $r=r_-$ in a finite affine parameter, but they also 
approach arbitrarily close to the ``top left" portion of the
inner horizon (which is also at $r=r_-$ and {\it is} included in the 
Eddington-Finkelstein patch) in the sense described above.}

The difference in the behavior of the outgoing light rays inside
the black hole is crucial for the fates of both the matter that
excites the extremal black hole and the partners of the Hawking 
radiation. Basically it means that all the information ends
up just inside the event horizon. To explain
the idea let me first consider the
Hawking process in a cartoon using the geometric optics
approximation and purely radial light rays (cf. Fig. \ref{decayfig}). 
Later I discuss the modifications of this picture brought in
by angular momentum and wave behavior. These modifications do 
not change the essence of the picture for the Hawking radiation,
but they are required to understand how the information in the
injected mass reapproaches the horizon from the inside. 

\subsection{Fate of the Hawking partners}

The pair to be created in the Hawking process
straddles the event horizon\footnote{The validity of this 
localized pair creation picture 
of the Hawking process has been elucidated by the work of  
and Parentani and Brout\cite{PareBrou}.}; 
equivalently, since the event horizon 
and apparent horizon are so close together (cf. (\ref{epsilon})),
the pair straddles the apparent horizon. While the Hawking quantum
on the outside escapes, the partner falls across the trapped
region and asymptotically approaches the inner horizon. As the
black hole decays the inner horizon moves outward toward
$r=Q$ on a {\it spacelike} trajectory as discussed in section
\ref{model}. The partner must stay
on a null ray, so it {\it crosses} the inner horizon into 
the interior region with $f>0$. At this point it begins moving outward
and asymptotically approaches the surface $r=Q$. 

Evidently, then, all the partners of the Hawking radiation pile
up just inside the event horizon, rather then falling into 
the black hole. This means that the information in their correlations
with the Hawking quanta is available just across the horizon. 
For a neutral evaporating black hole,  one can always choose
to foliate the spacetime with surfaces that dip way back to
the past in order to intercept all the partners before they
reach the singularity. However this is an extremely distorted
surface and, more importantly, once the black hole evaporates
completely this possibility is no longer available. By contrast,
in the near-extremal case, there is no need to dip the surface
at all, and  the information can be accessed in this way
at any time, even long after the hole has decayed entirely
back to extremality.

\subsection{Horizon instability}
Now let us consider more closely the role of the back reaction
in the decay process.
The partners carry negative energy. According to the preceding 
discussion, this energy
piles up with ever increasing density just inside
the event horizon. When the back-reaction is 
accounted for, this leads to some kind of instability 
of the horizon. 

In arriving at this conclusion I have assumed that 
the partners do not encounter a region of large stress energy
(where the back-reaction would be large and the adiabatic approximation 
would break down) {\it before} returning to the event horizon.  
However there seems to be a chance that 
exactly this would happen as the inner apparent horizon is approached.
In a static, non-extremal Reissner-Nordstr\"om metric, the negative
energy partners are stuck in the ergoregion between the two horizons
and they pile up just outside the inner apparent horizon. In the
Unruh vacuum, an observer who freely falls across this horizon
at late times sees a very large outgoing negative energy 
density which grows exponentially with the time at which the
observer falls across the inner apparent horizon.\footnote{Contrary 
to what I said in an earlier draft of this
paper, this is {\it not} the same as the quantum
instability at the {\it Cauchy} 
horizon of stationary non-extremal charged 
two-dimensional\cite{HiscInst2d}, 
four-dimensional\cite{BirrDavi}, and charged rotating four-dimensional
\cite{HiscInst4d} black holes. On the Cauchy horizon the stress-energy
tensor is actually {\it infinite}, and an observer who falls across the
Cauchy horizon sees this divergence.}
Although this growing flux is purely outgoing, there is also presumably
some finite ingoing flux, and together these mean that the invariant
square of the stress tensor is getting large so there should be a large
back-reaction. 
Might the same instability occur near 
the inner horizon of the {\it decaying} near-extremal black hole?
After all, this black hole remains non-extremal
for a very long time of order $Q^4/\mu$ (\ref{halflife}),
so perhaps there is a buildup of negative energy just outside the
inner horizon in this case as well. 
If so, then rather than crossing the inner horizon and 
finally returning to the event horizon, the partners may encounter
a (spacelike or null) singularity and never make it across the inner 
horizon and back out to the event horizon. 

It seems however that such an instability of the inner horizon 
probably does {\it not} occur. The negative energy
drives the inner horizon out on a spacelike trajectory towards the
event horizon, which allows the negative energy to slip 
across before piling up too much. 
I have made a rough estimate of the amount of negative energy density
that builds up, and it turns out to be of order $\mu/Q(\ln Q)^2$ which is
very small (compared to Planck density)
for the near-extremal black holes we are considering. 
The calculations behind this estimate are given in the Appendix of this
paper. Though not rigorous, this argument at least makes the assumption
the inner horizon is {\it not} unstable fairly plausible. 
It would be very interesting to check the validity of this assumption
in a self-consistent numerical calculation like that of Ref. \cite{LoweO'Lo}.
In the discussion section I will return to the instability
issue.\footnote{A study of the stability of the interior of an evaporating
Reissner-Nordstr\"om black hole was made by Kaminaga\cite{Kami}, who
used a two-dimensional model so that the stress tensor could be 
computed exactly. In this work it was assumed that both the mass and
the charge evaporate, with a fixed ratio, and the stress tensor was 
computed on this evaporating background. No attempt was made to enforce
self consistency in the sense of (say) the four-dimensional semi-classical
Einstein equation. One of Kaminaga's results was that the stress tensor
blows up at the {\it Cauchy} horizon that forms. Unfortunately this study is
of no direct help to us, since we are interested in decay 
with the charge fixed. This makes a big difference, because for us the 
inner horizon moves {\it out} whereas in the model of 
\cite{Kami} it moves {\it in}.
Also, the lack of self-consistency could well be important to the question of
stability.} 
 
\section{Angular momentum and wave behavior}

The restriction to radial light rays was convenient in
the preceding discussion. However, both the injected matter
and the Hawking radiation may carry angular momentum, and 
they may scatter as waves, so
it is necessary to see whether the basic picture described
in the previous section survives when these effects are included.
The Hawking radiation in anything other than the $S$-wave is
suppressed 
by a factor $(\omega Q)^l$ which is a very small number 
of order $(\mu/Q)^{l/2}\ll 1$ for $\omega$ of order the 
Hawking temperature (cf. the surface gravity (\ref{kappa})). Nevertheless,
it is still important in principle to determine the fate of the 
partners with $l\ne0$, and moreover the injected matter need not have $l=0$.

Let me begin by sticking with the geometric optics approximation.
The conclusion will be that, except for the ingoing radial null
geodesics, all the timelike or null geodesics miss the singularity
and swing back out to asymptotically approach the event horizon.
I will then extend the analysis to wave propagation, arriving
at two conclusions. First, for the negative energy 
wavepackets inside the horizon
there is very little scattering, and second, there
is a reasonable way to handle the singularity, for which waves simply
scatter back to large radius without being swallowed.

\subsection{Geodesics with angular momentum} 
The geodesics in the line element (\ref{Vaidya}) with constant
$Q$ and $M$ may be taken to lie in the plane $\theta=\pi/2$.
The remaining coordinates are functions of an affine parameter
$\lambda$ (which we take to be the proper time in the timelike
case),
$(v(\lambda), r(\lambda), \phi(\lambda))$. Defining the
conserved energy $E=g_{v\alpha}\dot{x}^\alpha=f\dot{v}-\dot{r}$ 
and angular momentum $L=g_{\phi\alpha}\dot{x}^\alpha=r^2\dot{\phi}$,
the equation 
$g_{\alpha\beta}\dot{x}^\alpha\dot{x}^\beta=\eta$
($\eta=0$ for null curves and $\eta=1$ for timelike curves)
becomes an equation for the radial coordinate:
\beq
\dot{r}^2 + U(r) = E^2
\label{radeqn}
\eeq
with the effective potential $U(r)$ given by 
\beq
U(r)=f(r)(L^2/r^2 + \eta).
\label{U}
\eeq
 
Near $r=0$ we have $f(r)\simeq Q^2/r^2$, so there is a tremendous
barrier unless both $L$ and $\eta$ vanish. That is, only radial
null geodesics can reach the timelike singularity. Any other 
geodesic with $\dot{r}<0$ inside the horizon reaches a
minimum value of $r$ and then proceeds to increasing values of 
$r$. The advanced time $v$ goes to infinity when the inner horizon
is crossed with $\dot{r}>0$ for positive energy orbits and
with $\dot{r}<0$ for negative energy orbits (since 
$\dot{v}=(E+\dot{r})/f$). In the extremal case, for which
$f(r)=(1-Q/r)^2$, there are no negative energy orbits, and the geodesics
all return to $r=Q$, crossing the inner horizon at a finite affine 
parameter as described previously for the  radial, massless case.
As in the radial case, these geodesics also come arbitrarily close
to the event horizon. 

The actual situation in the case of the decaying near-extremal
black hole is of course time-dependent, but we can understand
the nature of the trajectories there by considering an initial
segment, propagating in a static, near-extremal, Reissner-Nordstr\"om
metric, followed by a second segment beginning after the trajectory
reaches its minimum value of $r$.

On the initial segment of a positive energy orbit (which is
the relevant type of orbit for the infalling matter), 
the inner horizon at $r_-$ is crossed after a 
finite affine parameter and a finite advanced time $v$. After reaching
its minimum value of $r$ at finite $v$ the geodesic begins to move
to larger $r$, reaching $r_-$ again at finite affine parameter
but infinite $v$. That is, the geodesic asymptotically approaches
$r_-$ from the inside. In the decaying case, the inner horizon
(the inner zero of $f$) gradually moves outward toward the event 
horizon. Our geodesic follows it out, just as in the purely radial
case described earlier. 

The negative energy orbits
(which are the relevant ones for the Hawking partners) behave a bit  
differently. In the static phase they take an infinite
advanced time to cross the inner horizon on the way in. 
In the decaying case, they slip across the inner horizon
after a finite advanced time. At the inner horizon they have
$\dot{r}=-(L^2/r^2 +\eta)/2\dot{v}<0$. Since they spend
a long time near the almost static inner horizon, this
negative $\dot{r}$ must be very small. The orbits thus continue
in a little bit to some minimum radius, and then move out again 
following the inner horizon to the event horizon.
Thus, in fact, allowing for angular momentum
changes nothing essential about the process.
 
\subsection{Wave propagation}

The geometrical optics analysis suggests that if the quantum
field is treated properly using a wave equation, the singularity
may be avoided and the waves will scatter back out to the
horizon. To study this question, let us 
consider for example matter satisfying the massless
Klein-Gordon equation. The scattering can
cause dispersion of wavepackets, so the simple picture of where the
Hawking partners go may need revision.
Even the outgoing  
$S$-wave just inside the horizon can in principle backscatter from the 
curvature and fall in towards the origin. More fundamentally, 
the timelike singularity at $r=0$ must be tamed in some fashion
in order to make sense of the wave equation. This amounts to 
the question of boundary conditions at the origin.

To understand the basic physics, it should be adequate to
consider only the static near-extremal Reissner-Nordstr\"om
spacetime. The wave equation
can be separated in both Eddington-Finkelstein type coordinates
(\ref{Vaidya}) and diagonal coordinates 
\beq
ds^2=f (-dt^2 + dr_*^2) + r^2(d\theta^2+\sin^2\theta d\phi^2)
\eeq
where 
\beq 
f(r)= (r-r_-)(r-r_+)/r^2
\eeq
and the tortoise coordinate $r_*$ is related to $r$ by
\beq 
dr_*=f^{-1} dr.
\label{tortoise}
\eeq
The diagonal coordinates cover the entire interior up to $v=\infty$
in the extremal case, but they are singular on the inner horizon
in the non-extremal case. Thus one can use them to discuss either
the region between the two horizons or the region inside the
inner horizon. (To discuss both regions together the
Eddington-Finkelstein coordinates would be preferred.)  

Writing the matter field $\Phi$ as
\beq
\Phi=\frac{u(r)}{r}Y_{lm}(\theta,\phi)e^{-i\omega t}
\eeq
the wave equation reduces to the radial equation
\beq
-\frac{d^2}{dr_*^2}\, u=[\omega^2-V(r)]\, u
\label{radeq}
\eeq
with 
\beq
V(r)=f\, [f'/r+l(l+1)/r^2]
\label{V}
\eeq
where $f'=df/dr$.
Note that the second term in $V(r)$ is similar to the effective
potential (\ref{U}) for massless geodesic motion, but there is 
no geodesic analog for the $f'$ term.

The geometrical optics limit of the negative energy wavepackets
is described by negative energy light rays.
Let us follow the trajectory of these light
rays and estimate the amount of scattering suffered by a wavepacket.
The rays that begin just inside the horizon $r_+$ sink down to $r_-$
asymptotically approaching the inner horizon as discussed in the
previous subsection. In between
$r_-$ and $r_+$ the maximum value of $|ff'|/r$ is
bounded by
\beq
 |ff'|/r < |f_{\rm max} f'(r_-)|/r_-=(r_+-r_-)^3/4r_+r_-^4
\sim\mu^{3/2}Q^{-7/2}.
\eeq
For $S$-waves we therefore have 
$\omega^2/|V|\gsim(\omega/T_H)^2(Q/\mu)^{1/2}$,
so there is extremely little scattering except possibly for very low
frequencies compared to the Hawking temperature.

For nonzero angular momentum, although the potential term in 
(\ref{radeq}) is important, there 
is still very little wave scattering, since the wavelength remains small
compared to the radius of curvature $Q$ of the spacetime. To see
this note that, when $l\ne 0$, 
$V(r)$ is negative everywhere between $r_-$ and $r_+$. 
Thus in the eikonal approximation we have $k_*>\omega$.
Since $k_*=fk$ (\ref{tortoise}), this yields 
\beq
k>\omega/f>\omega/|f|_{\rm max}\gsim (\omega/T_H)(\mu Q)^{-1/2}
\eeq
(which holds also for $S$-waves).
Thus, although the wavelength does not remain small compared with 
$r_+-r_-\sim\sqrt{\mu Q}$, it certainly does remain small compared 
to the radius of curvature $Q$ except for very low frequencies
$\omega\lsim (\mu/Q)^{1/2} T_H$.
 
The negative energy wavepackets are thus well described by 
the geometric optics approximation when they cross the inner horizon, 
and it is clear the approximation will remain good the rest of 
the way. As the metric approaches the extremal one, the 
trajectory proceeds outward towards the event horizon,
always remaining at small values of $f$.

The positive energy wavepackets on the other hand will clearly
fall deep into the black hole and scatter. In the geometrical
optics approximation we saw that trajectories with angular 
momentum will dip inside $r_-$ before coming up again. I will
not attempt here to estimate the amount deviation from geometric
optics, but rather ask what is the general nature of scattering
close to the origin.  As long as these modes do not get
swallowed or trapped by the  singularity, it is quite plausible
that they too eventually make it  back out to the event horizon.
Thus let us  next consider scattering deep inside the black hole
near the singularity.  

To this end note that
the most singular term in (\ref{V}) is not the centrifugal
barrier but the $f'$-term, which diverges as $-2Q^2/r^3$ near
$r=0$. Meanwhile $f$ diverges as $Q^2/r^2$, so we have
\beq
V(r) \simeq -\frac{2Q^4}{r^6}\qquad \mbox{as}\quad r\rightarrow 0.
\eeq
To use the radial equation (\ref{radeq}) this should
be expressed in terms of $r_*=\int f^{-1}\, dr \simeq r^3/3Q^2$,
yielding
\beq
V(r_*) \simeq -\frac{2}{9}\frac{1}{r_*^2}.
\eeq

The Schr\"odinger-like equation (\ref{radeq}) 
in an attractive potential  
$-\gamma/r_*^2$ has solutions near the origin 
of the form $r_*^s$, where $s=(1\pm\sqrt{1-4\gamma})/2$.
If $\gamma>1/4$ the solutions oscillate an infinite number
of times as the origin is approached. 
In our problem $\gamma=2/9$, so there are no oscillations
and we have $s=2/3$ and $s=1/3$. 
The behavior of $u(r)/r$ for these two solutions is 
$r^1$ and $r^0$. Thus there is no ``singular" behavior at 
the singularity. On the other hand, since both solutions 
are square integrable in the appropriate measure\footnote{This is the
measure for which the spatial differential operator in the 
wave equation is symmetric\cite{HoroMaroQP}, i.e., $f^{-1/2}dv$, 
where $dv$ is the proper
volume element.}, a boundary 
condition must be supplied to select a unique solution. 
This boundary condition is presumably supplied by the physics
at the singularity. The phase shift between incoming and 
outgoing waves depends on this boundary condition, but the
fact that the ingoing partial waves emerge as outgoing waves 
and return to large values of $r$ does {\it not}.
For our purposes this is enough.\footnote{The existence of the
two regular solutions means that the Hamiltonian is not
``essentially self-adjoint". In \cite{HoroMaroQP} Horowitz and 
Marolf investigated when the
Hamiltonian for a Klein-Gordon field remains essentially 
self-adjoint in the presence of timelike singularities.
They mention the result found here for the Reissner-Norstr\"om
black hole, but also identify other black hole singularities for which  
essential self-adjointness does hold.}

\section{Discussion}
\label{disc}

Now that we have some insight into the semiclassical
decay of near-extremal black holes let us compare
it with the corresponding D-brane decay. A natural
process to consider is the excitation of an extremal
configuration followed by decay back to the extremal
state. A complication arises, however, in the D-brane
description of this process. Even if the extremal D-brane
configuration
is in the maximal entropy state (the microcanonical
ensemble for the fixed set of charges) to begin with
(which is the usual assumption in D-brane calculations\footnote{The
mixed character of these D-brane states was recently emphasized
by Myers\cite{MyersPure}.}),
after being excited its state will depend in a complicated
way on the state of the absorbed quanta and its interaction
with them. Since the agreement found between D-brane and black
hole entropies and radiation rates holds when the D-brane
state has maximal entropy, the two processes would
already differ immediately after the energy absorption.
Thus instead let us consider a process that starts off with
an already excited near-extremal D-brane state 
with maximal entropy, and compare this to the semiclassical
decay of a near-extremal black hole.

The string dynamics is unitary, so the final state has
the same entropy as the initial one. This state is correlated
in the D-brane and radiation degrees of freedom. The reduced 
density matrix of the D-brane configuration is
the maximal entropy state (since the original
state before decay was the microcanonical ensemble), and the
reduced density matrix of the radiation is that of thermal
radiation from a ``greybody". 
The semiclassical decay is also unitary in the standard sense
of quantum field theory in curved spacetime if
the background is treated classically and if the horizon
instability is ignored. Then
the state of the quantum field itself remains pure if it started
out pure, with the Hawking radiation
correlated to the state of the field inside the 
horizon.\footnote{Allowing the background spacetime to 
fluctuate would go beyond
the semiclassical calculation, and would presumably also preserve
purity of the joint state.} 

Although the spacelike surfaces
that foliate the extremal black hole spacetime include the
timelike singularity at the origin (which has no counterpart
in the D-brane configuration), we have seen that the  
Hawking partners essentially never go near the singularity
but rather end up hovering just inside the event horizon,
and the waves that do scatter deep into the black hole just
scatter back out again and approach the horizon (although
the phase shift upon scattering through the origin depends
on an unknown boundary condition there). Thus the singularity
at $r=0$ appears to be irrelevant to the unitarity question
in this case.

Neglecting the horizon instability is, however, inconsistent
in the semiclassical approach, in which the black hole
background decays in response to quantum field energy.
What is the nature of this instability? I do not know, but 
I will offer some speculative remarks.
If the inner apparent horizon is indeed stable, as argued 
in the Appendix, then there is certainly pile up of negative energy
Hawking partners (as well any previously injected positive 
energy\footnote{The positive energy
of the injected matter should be balanced by the total negative
energy flux into the black hole, and if these two energies were
identically distributed, one might expect them to cancel,
leaving the horizon stable. However there is no reason why the
injected matter and the Hawking radiation partners should end
up equally distributed in general. 
Even if the injected matter is purely an $S$-wave, when it
arrives back at the horizon it will be 
located at slightly smaller radius than all the Hawking 
partners (which are also primarily $S$-waves) 
inside the event horizon. This would produce a ``dipole" structure in the
energy density.}) just behind the inner apparent horizon
which at late times is just inside the event horizon.
This is a purely null flux, however, so it will yield a large
invariant such as $T_{\mu\nu}T^{\mu\nu}$ only if there is also
a finite ingoing flux. There is of course the ingoing
negative energy flux associated with the Hawking radiation,
but this is going to zero at late times. I do not know 
what the late time limit of this invariant is. A large
invariant will certainly develop if more matter is injected
to re-excite the black hole above extremality. In this case,
the instability would be located just inside where the event
horizon would have been had the new matter not been injected.
The actual event horizon, on the other hand, moves out in anticipation
of the arrival of the injected matter, so if enough matter is
injected the event horizon will be far outside this instability! 

For the purposes of comparison with the D-brane picture,
the most important question about the buildup of  
energy inside the horizon is whether or not it affects the 
subsequent Hawking radiation in any way. If it were located
where the Hawking pairs are born, it would certainly 
have an impact. However, after the extremal hole is re-excited, 
the event horizon has moved away from the region of large energy density,
and when they are born the new Hawking pairs do not see this large 
energy density. As usual, 
to determine the pair creation amplitude one must follow the
pair backwards in time to see if it emerges from a vacuum state.
In this case, even though the pair gets very close to the
inner apparent horizon during the quiescent period when the
hole is extremal, the pair never gets {\it inside} the apparent horizon,
so even in its past it never seems to see the large energy density.
This naive argument thus suggests that the subsequent Hawking
radiation is insensitive to presence of the large energy density
provided the event horizon itself is indeed not singular.

However, even if the event horizon is not {\it necessarily} singular
after the near-extremal black hole initially decays, the huge
null negative energy flux behind the horizon is bizarre and 
makes the situation highly unstable, since the slightest influx
would be catastrophic. 
It seems at first that such an unstable event horizon would
ruin the agreement with the D-brane picture, but in fact
there is also evidence from the theory of D-brane decay that  
a horizon instability occurs. 
When a near-extremal,
nonsingular (in the sense of the strong coupling analog) 
D-brane state decays, the final state
is a mixed and the Hawking radiation is correlated to the
microstate of the D-brane configuration.
In any particular realization of the Hawking radiation, 
a partial projection onto a sub-ensemble
with a {\it non-uniform} distribution of charge presumably 
occurs. 
Small changes of the internal charge 
distribution of a D-brane state whose strong coupling analog
is a black hole (or string) with nonsingular horizon seem 
to generally  
correspond to a black holes with a singular ``would-be'' 
horizon\cite{HoroMaroSing, KaloMyerRous, HoroYang}. 
It therefore appears that the D-brane configuration 
evolves into a state whose strong coupling analog has a singular
horizon.\footnote{A partial projection would also occur 
in the {\it pure} state semi-classical description of the Hawking radiation.
In that case Massar and Parentani
\cite{MassPare} showed that a quantity that may be called the
conditional expectation value of the stress tensor is huge and 
oscillating near the horizon. However, in this pure state case,
blurring the ``post-selection" of the state can presumably
average over the wild fluctuations leaving a small residual.
In the mixed state D-brane case, {\it every} member of the 
ensemble is ``singular", and one of them  
is in principle {\it actual}.}

A horizon instability would mean a failure of the semiclassical
approximation, which may be just what is needed to avoid a discrepancy
with the D-brane picture over the question of entropy of the reduced
state of the radiation 
 when a black hole or D-brane configuration is repeatedly excited 
and allowed to decay. In the D-brane case, the entropy of the
final D-brane configuration itself is bounded by the maximal entropy
of the microcanonical ensemble. Since the whole process is unitary,
the entropy of the radiation cannot just be that of so much
thermal radiation. Instead, there must be correlations in the
radiation emitted at different times. In the black hole case,
if the horizon remains regular, Hawking's analysis requires
that the radiation be purely thermal. This would be consistent
with unitary evolution in the semiclassical framework once
the degrees of freedom of the quantum field inside the black hole 
are accounted for. However, the reduced state of the  Hawking
radiation would have a much larger entropy than that
of the D-brane radiation. If instead the horizon is unstable after
the initial decay process, then perhaps the repeated excitation of the
black hole cannot be described in the semiclassical approximation.

The preceding discussion seems at first to offer a possible
resolution of how it could be that the D-brane and semiclassical
black hole calculations agree so well with respect to the rate
of radiation, yet disagree as to the long term evolution---although
we have to destroy the black hole in order to save it.
However, it seems this cannot be the whole answer, since a small
change in the process leads to trouble. Instead of repeatedly
exciting the system and letting it decay, one might send in a 
constant flow of positive energy matter to maintain the system
indefinitely in a near-extremal state. In this case the Hawking
partners pile up at the inner horizon and stay there, well away 
from the event horizon. It is the inner horizon that is then unstable,
but this does not influence the Hawking radiation. Disagreement
with the D-brane picture over the state of the radiation then seems
unavoidable. Another example of a process where the pictures disagree
was disussed by Maldacena and Strominger\cite{MaldStro}. 
They pointed out that one can find a regime in which it is
possible to excite an extremal black hole to a near-extremal
configuration whose corresponding black hole entropy is much greater 
than the entropy of the original extremal configuration. 
In such a process the resulting Hawking radiation would have
a much larger entropy than the corresponding D-brane could radiate.

In view of these examples, it is clear that something is still
missing in our understanding of the uncanny (partial) 
agreement between
the weakly coupled D-brane model and the strongly coupled
black hole model. Hopefully the semiclassical study presented
here can be helpful in addressing this question. 

\section*{Acknowledgments}

I am grateful to R.C. Myers for many helpful discussions,
suggestions, and criticism. I would also like to thank
D. Brill, S. Corley, S. Droz, V. Frolov, J. Louko, J. Maldacena, 
D. Marolf, A. Ori, D. Page, R. Parentani, and E. Poisson for helpful 
discussions. This work was supported in part by NSF grant 
PHY94-13253. 

\section*{Appendix: Energy density at the inner apparent horizon}
For a static Reissner-Nordstr\"om black hole the inner (apparent)
horizon is an infinite blue shift surface.
Between the inner and outer horizons an outgoing light
ray falls inward and asymptotically approaches the inner  
horizon. An outgoing null flux of (negative) energy between the horizons 
will thus pile up just outside the inner horizon. In the static
case, the energy density observed on a free-fall world
line crossing the inner horizon will grow exponentially like 
\beq
\rho\sim e^{2\kappa v} \rho_0,
\label{rho}
\eeq
where $\kappa$ is the surface gravity 
of the inner horizon, $v$ is the advanced time, and $\rho_0$ is some
initial energy density. This quickly exceeds the Planck density
so, as long as there is a finite ingoing flux as well, the square
of the stress tensor will be a large invariant and the back-reaction
will be large. 

In the evaporating case the inner horizon
is moving out, allowing the energy to slip across. Here we estimate
how large the energy density gets near the inner horizon. 
We do this by estimating how much advanced time passes before a 
given outgoing light ray crosses the inner horizon.

In the static near-extremal case, an outgoing
light ray it falls from the midpoint 
$(r_+ + r_-)/2$ to the position $r_- + x$ in an advanced time $v$ 
given by 
\beq
x\sim (r_+ - r_-)\, e^{-\kappa v} \sim \kappa Q^2\, e^{-\kappa v},
\eeq
where $\kappa$ is the surface gravity (\ref{kappa}) which is
approximately the same for both horizons. If the hole is 
evaporating slowly this should still hold to a good approximation.
Meanwhile the inner horizon is moving out. At the inner horizon
$f(v,r_-(v))=0$, so 
$\dot{r}_-=-f_v/f_r\simeq -\dot{\mu}/\kappa Q\sim \kappa^3 Q$.
The change $\Delta r_-$
in the radius of the inner horizon over a time $v$ is approximately
$\dot{r}_- v$, from which we find
\beq
\Delta r_- \simeq  \kappa^3Qv.
\eeq
The time at which the outgoing light ray meets the inner horizon
is now found by setting $\Delta r_- = x$, which yields
\beq
e^{\kappa v}\sim Q/\kappa^2 v.
\eeq
Since $Q/\kappa\gg1$, we find 
\beq
e^{\kappa v} \sim Q/\kappa\ln Q.
\eeq
Using this and choosing the initial density $\rho_0\sim \kappa^4$
(which seems reasonable for the Unruh vacuum)
we obtain from (\ref{rho}) 
\beq
\rho_{\rm max}\sim \frac{\mu}{Q(\ln Q)^2}\ll 1.
\eeq
By this estimate the energy density stays very small at the
inner horizon of an evaporating near-extremal black hole.

\end{document}